\documentstyle[twoside,fleqn,espcrc2,epsf,amssymb]{article}

\newcommand{\di}{\mbox{$\not\!\!D$}}
\newcommand{\rhb}{\mbox{$\overline{\rho}$}}
\newcommand{\sib}{\mbox{$\overline{\psi}$}}
\newcommand{\ib}{\mbox{$\overline{\rm I}$}}
\newcommand{\ch}{$\chi$SB}

\title{Chiral symmetry breaking, instantons and the ultimate quenched calculation.}

\author{U. Sharan\address{Department of Theoretical Physics,
University of Oxford.} and M. Teper\address{All Souls College and
Department of Theoretical Physics, University of Oxford,\\ 1 Keble
Road, Oxford, OX1 3NP, United Kingdom.}}
       
\begin{document}

\begin{abstract}
We calculate the spectral density of the Dirac operator over an
ensemble of configurations composed of overlapping instantons and
anti-instantons. We find evidence that the spectral density diverges
in the limit $\lambda \rightarrow 0$. This indicates the breaking of
chiral symmetry and also provides evidence that quenched QCD may be
pathological in nature.
\end{abstract}

\maketitle

\section{Introduction}
The reader is referred to \cite{Col,Shu} for background to this
field. Preliminary results for this work have been given elsewhere
\cite{Tep}.

The QCD Lagrangian for $N_{f}$ quarks in the chiral limit possesses a
large family $SU_{L}(N_{f}) \otimes SU_{R}(N_{f}) \otimes U_{B}(1)
\otimes U_{A}(1)$ of global symmetries. Phenomenologically we find
that the non U(1) part of this symmetry is dynamically broken to the
diagonal subgroup, $SU_{L}(N_{f}) \otimes SU_{R}(N_{f}) \rightarrow
SU_{D}(N_{f})$. This chiral symmetry breaking (\ch) overcomes
the degeneracy between parity partners (particles with opposite
parities but otherwise identical quantum numbers) and gives rise to
$N_{f}^{2} - 1$ Goldstone bosons (a role played by the 3 pions for
$N_{f} = 2$). The order parameter for \ch\ is given by:
\begin{equation}
\label{chiord}
\langle\sib\psi\rangle = \lim_{m \rightarrow 0}
i\int_{0}^{\infty}\frac{2m\rhb(\lambda,m)}{\lambda^{2} + m^{2}}
\end{equation}
where \rhb$(\lambda,m) = \lim_{V \rightarrow
\infty}V^{-1}\langle\rho(\lambda)\rangle_{\rm m}$. We note that the spectral
density \rhb$(\lambda,$m) of the Dirac operator explicitly depends
upon the mass m via the fermion determinant. Conventionally the above
integral is calculated in the appropriate limits to give
\begin{equation}
\label{bc}
\langle\sib\psi\rangle = i\pi\rhb(0)
\end{equation}
This however can break down if the spectral density is not smooth in
m. To illustrate, \rhb\ $= \lambda^{-\alpha}, \alpha \in (0,1)$
implies $\langle\sib\psi\rangle = i\pi m^{-\alpha}$, hence the
order parameter diverges in the chiral limit.  If however \rhb\
$=(m/\lambda)^{\alpha}$ then we have a perfectly good order parameter
in spite of the spectral density diverging in the limit of small
eigenvalues. Our first conclusion is therefore that one should use
(\ref{chiord}) to calculate the chiral condensate at various masses
and then extrapolate to the chiral limit. One can however rely on the
Banks-Casher relationship (\ref{bc}) for quenched calculations, for
here there is no fermion determinant to consider.

\section{Calculation of spectral density}

We construct an explicit matrix representation of the Dirac operator
i\di[A] for a given gauge field A consisting of $n_{-}$ instantons (I)
and $n_{+}$ anti-instantons (\ib) (each object in isolation has gauge
field $\widetilde{A}_{i}^{\pm}$). The matrix is of course defined by
$(i\di[A])|j\rangle = D_{kj}|k\rangle$ in terms of some basis \{$|\rm
i\rangle$\}. If the basis is orthonormal then $D_{kj} = \langle
k|i\di|j\rangle$. The basis we choose to use is the basis of ``would
be zero-modes''
\{${|\psi_{1}^{+}\rangle,\ldots,|\psi_{n_{+}}^{+}\rangle,|\psi_{1}^{-}\rangle,\ldots,|\psi_{n_{-}}^{-}\rangle}$\}
where $(i\di[\widetilde{A}_{i}^{\pm}])|\psi_{i}^{\pm}\rangle = 0$. We
know that:
\begin{eqnarray}
\langle\psi_{i}^{+}|i\di|\psi_{j}^{+}\rangle & = & 0\nonumber\\
\langle\psi_{i}^{-}|i\di|\psi_{j}^{-}\rangle & = & 0\nonumber\\
\langle\psi_{i}^{+}|i\di|\psi_{j}^{-}\rangle & \doteq &
V_{ij} =V(x_{i}^{+},\rho_{i}^{+},x_{j}^{-},\rho_{j}^{-})
\end{eqnarray}
The first two follow from $\gamma_{5}|\psi_{i}^{\pm}\rangle =
\pm|\psi_{i}^{\pm}\rangle$ and $\{\gamma_{5},i\di\} = 0$. The function
V (which is known approximately for a single I-\ib\ pair) in the third
equation depends upon the center and size of the two instantons. In
general it should also depend upon their relative colour orientation
but we choose to ignore this, this should be irrelevant as long as
there are no non-trivial correlation effects in colour space. In terms
of this basis we can construct the $(n_{+}+n_{-})\times(n_{+}+n_{-})$
matrix $D \equiv \langle\psi|i\di|\psi\rangle$ which has block
zeroes on the diagonal and $V, V^{\dagger}$ respectively off block
diagonal.

There are two main objections to the matrix D being a matrix
representation of the Dirac operator. The first is that the ``basis''
we have chosen does not span the Hilbert space of wavefunctions. This
is not too important for our considerations as wish to study the
spectral density at small eigenvalues, these are the eigenvalues which
have split from zero due to instanton interactions. The second more
fundamental objection is that the ``basis'' we have chosen is not
orthonormal
\begin{eqnarray}
\langle\psi_{i}^{+}|\psi_{j}^{+}\rangle & = & U(x_{i}^{+},\rho_{i}^{+},x_{j}^{+},\rho_{j}^{+})\nonumber\\
\langle\psi_{i}^{-}|\psi_{j}^{-}\rangle & = & U(x_{i}^{-},\rho_{i}^{-},x_{j}^{-},\rho_{j}^{-})\nonumber\\
\langle\psi_{i}^{\pm}|\psi_{j}^{\mp}\rangle & = & 0
\end{eqnarray}
hence the matrix D is simply not a representation. We ignore the
colour orientation of U for the reason given in the consideration of
V. The third equation equals zero again due to the $\gamma_{5}$
structure. We can however construct a new orthonormal basis
$\{|\widetilde{\psi}^{+}\rangle,|\widetilde{\psi}^{-}\rangle\}$ using
the standard Gram-Schmidt procedure.
\begin{eqnarray}
|\psi_{j}^{+}\rangle & = & R_{ij}|\widetilde{\psi}_{i}^{+}\rangle\quad 1 \leq
j \leq i \leq n_{+}\nonumber\\
|\psi_{j}^{-}\rangle & = & S_{ij}|\widetilde{\psi}_{i}^{-}\rangle\quad 1 \leq
j \leq i \leq n_{-}
\end{eqnarray}

We are finally in a position to construct a matrix representation of the
Dirac operator:

\[
i\di \doteq \widetilde{D} \equiv
\langle\widetilde{\psi}|i\di|\widetilde{\psi}\rangle
\]

\begin{equation}
\rm{where}\ \ \widetilde{D}_{ij} = \left(\begin{array}{cc}
0 & \widetilde{V} = (R^{-1})^{\dagger} VS^{-1} \\
\widetilde{V}^{\dagger} & 0 \\
\end{array}\right)
\end{equation}

This matrix representation shares many of the properties of the Dirac
operator (as it must). Firstly the matrix is endowed with a rich
structure, the elements are not random. This must be so as vectors in
a Hilbert space obey triangle inequalities (a fact which is crucial is
attempting to orthogonalize the basis). Secondly, non-zero eigenvalues
occur in pairs $\pm\lambda$ hence the matrix satisfies the $\gamma_{5}$
symmetry. Thirdly the Atiyah-Singer theorem is obeyed, a gauge field
configuration with net winding number $Q = |n_{-} - n_{+}|$ does
indeed have (at least) Q exact zero eigenvalues. Fourthly our
representation captures the essence of the mechanism by which
eigenvalues are split away from zero. To see this simply consider the
case of an I-\ib\ pair and notice that if we choose the
function V appropriately then we will get the correct eigenvalue
splitting. All these properties give us hope that the matrix
representation indeed shares the essential qualities of the actual QCD
Dirac operator for determining the spectrum at small eigenvalues.

The only other thing to decide is which functions to use for U and V ?
Though these functions are known classically, there is little reason
to believe that the classical forms dominate in the QCD
vacuum. Pragmatically we use a variety of zero mode wavefunctions in
calculating U and V (hard spheres, Gaussian and classical zero mode)
and look for qualitative features of the resultant spectral densities
which are independent of the functions used. In this talk we simplify
and use the same functional form for both U and V, we discuss the
general case (amongst other things) in a subsequent publication
\cite{Te3}, but it suffices to say that the conclusions are
unaffected.

\begin{figure}[htb]
\leavevmode
\epsfxsize=70mm
\epsfbox{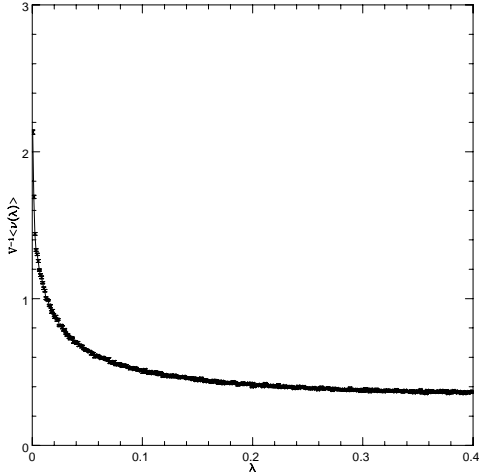}
\caption{Classical Zero Mode V = U, (${\mathbb R}^{4}$). Connected
line from V = 97 + $\overline{97}$. Points ($\times$) from V = 63 +
$\overline{63}$, 100000 configurations.}
\label{cl}
\end{figure}

\section{Results}

\begin{figure}[htb]
\leavevmode
\epsfxsize=70mm
\epsfbox{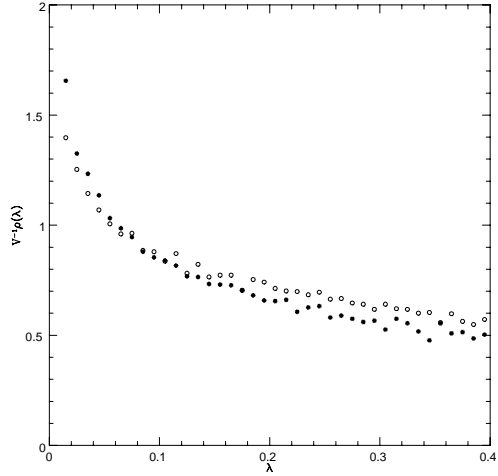}
\caption{Hard Sphere, (${\mathbb T}^{4}$). $\beta = 6.0,\ 32^{3}64,\ 51$
configurations. ($\circ$) Positions and sizes from lattice data, ($\star$)
positions random. Data from \cite{Te2}.}
\label{lt}
\end{figure}

Figure (\ref{cl}) shows the spectral density per unit volume for
$N_{f} = 0$ (quenched) for two different volumes. The instanton gas is
fairly dense, more dense than in the instanton liquid model
\cite{Shu} but less so than found in some recent investigations of
the vacuum \cite{Te2}. We have used the classical zero mode
wavefunctions for the overlap integrals. The most noticeable thing
about the resultant spectral density is the divergence at small
eigenvalues. This divergence is independent of the functions used, see
\cite{Te3} for a more in depth discussion. We can even use this
method to generate a spectral density from real lattice data (where
instanton positions and sizes are extracted as a study of the
vacuum). This is shown in figure (\ref{lt}).

If we increase the density of instantons then we na\"{\i}vely expect
greater interactions, leading to eigenvalues splitting away further
from zero. This should decrease the spectral density at small
eigenvalues. This is indeed seen, as is the converse. The fact that we
still see a divergence from the lattice data which is a dense gas
shows the divergence to be a generic feature. Intriguingly we also see
a difference if we position the objects at random, the spectral
density increases at small eigenvalues. This shows that there may be
non-trivial screening between instantons in the vacuum. When we
include the effect of the fermion determinant and simulate unquenched
QCD then we still find a divergence. This however results in a
perfectly acceptable quark condensate and $\langle Q^{2}\rangle(m)$
distribution in the chiral limit. Our work however indicates that
quenched QCD may be pathological in nature, it may have a divergent
quark condensate. If this is so, then why has such a feature not been
seen directly from the lattice ?  We believe that the spectral density
is protected by lattice artefacts, there are no small eigenvalues on
the lattice (unless by accident), hence the spectral density cannot
diverge as required. However such a feature should be apparent in the
``ultimate quenched calculation''.

\end{document}